\documentclass[useAMS,usenatbib]{mn2e}
\usepackage{epsfig}
\usepackage{multirow}
\usepackage{array}
\usepackage{color}

\newcommand{\jcp}{J. Chem. Phys.}
\newcommand{\aap}{A\&A}
\newcommand{\apj}{ApJ.}
\newcommand{\apjl}{ApJL}

\newcommand{\pdix}[1]{$\times$~10$^{#1}$}
\newcommand{\cmc}{cm$^{-3}$}

\newcommand{\mnras}{mnras.}

\title[Hyperfine excitation of N$_2$H$^+$ by H$_2$]{Hyperfine excitation of N$_2$H$^+$ by H$_2$: Toward a revision of  N$_2$H$^+$ abundance in cold molecular clouds}
\author[Lique et al.]{Fran\c{c}ois Lique$^{1,2}$\thanks{E-mail:
francois.lique@univ-lehavre.fr}, Fabien Daniel$^{3}$, Laurent Pagani$^{2}$ and Nicole Feautrier$^{2}$ \\
$^{1}$LOMC - UMR 6294, CNRS-Universit\'e du Havre, 25 rue Philippe Lebon, BP 1123, 76 063 Le Havre cedex, France \\
$^{2}$LERMA and UMR 8112, CNRS- Observatoire de Paris-Meudon, 5 Place Jules Janssen, 
 92195 Meudon Cedex, France\\
$^{3}$IPAG, Observatoire de Grenoble, Universit\'e Joseph Fourier, CNRS UMR5571, B.P. 53, 38041 Grenoble Cedex 09, France}
\begin{document}

\pagerange{\pageref{firstpage}--\pageref{lastpage}} \pubyear{2014}

\maketitle

\label{firstpage}

\begin{abstract}
The modelling of emission spectra of molecules seen in interstellar clouds requires the knowledge of collisional rate coefficients. Among the commonly observed species, N$_2$H$^+$ is of particular interest since it was shown to be a good probe of the physical conditions of cold molecular clouds.
Thus, we have calculated hyperfine--structure resolved excitation rate coefficients of N$_2$H$^+$(X$^1\Sigma^+$) by H$_2(j=0)$, the most abundant collisional partner in the cold interstellar medium. The calculations are based on a new potential energy surface, obtained from highly correlated {\it ab initio} calculations.
State-to-state rate coefficients between the first hyperfine levels were calculated, for temperatures ranging from 5~K to 70~K. 
By comparison with previously published N$_2$H$^+$--He rate coefficients, we found significant differences which cannot be reproduced by a simple scaling relationship.
As a first application, we also performed radiative transfer calculations, for physical conditions typical of cold molecular clouds. 
We found that the simulated line intensities significantly increase when using the new H$_2$ rate coefficients, by comparison with the predictions based on the He rate coefficients. In particular, we revisited the modelling of the N$_2$H$^+$ emission in the LDN 183 core, using the new 
collisional data, and found that all  three of the density, gas kinetic temperature and N$_2$H$^+$ abundance had to be revised. 
\end{abstract}


\section{Introduction}

The study of interstellar clouds requires to model the emission of various molecular species, either neutrals, ions, or radicals.
Such studies can aim at determining the molecular abundances, which put constraints on 
our understanding of the molecular formation chemistry. Additionally, various molecules can be used as probes of 
the physical conditions of the clouds (temperature, density, kinematics), and thus shed some light on the dynamical
evolution of the cores and their capability to form stars.
No single species is observable from the surface of the clouds to their deeply buried, dense and cold cores. As an exemple, while HF traces the outskirts \citep{Indriolo:2013cj}, CO originates from the bulk of the envelope but starts to be depleted at 
densities typically above $\sim$3 \pdix{4} \cmc\  (\citealt{Lemme:1995wz,Willacy:1998bl}; see \citealt{2012A&A...548L...4P} for a summary). Dense cores, and especially prestellar cores (density above 1\pdix{5} \cmc, \citealt{Caselli:2008dm}), suffer from strong
 depletion of most tracers (CO, CS, SO, etc., e.g., \citealt{2005A&A...429..181P}) and it has been found that only a handful 
 of species survive in the gas phase. Among them,
nitrogen--carrying molecules are the best known (NH$_3$, N$_2$H$^+$, and possibly CN, \citealt{Lee:2001jm,2002ApJ...569..815T,2007A&A...467..179P,2008A&A...480L...5H}). In particular, they have the advantage to be
 generally confined to the cores. 
 Among these species, N$_2$H$^+$ is an interesting tool to investigate cold cores \citep{2007A&A...467..179P}.
 In a first place, this is due to its high dipole moment \citep[$\mu=3.4 \pm 0.2$ debye,][]{Havenith:90}. Hence, contrary to the 23 GHz inversion lines of NH$_3$, its $J = 1-0$ line 
 critical density is high ($\geq$ 1 \pdix{5} \cmc, about 2 orders of magnitude higher) which makes this molecule a better tool 
 than NH$_3$ to evaluate the excitation conditions in prestellar cores. Additionally, its hyperfine structure strongly helps to discriminate between opacity effects and excitation temperature effects which puts strong constraints in the determination of the physical conditions. Finally, its deuterated
isotopologue was also shown to be valuable in quantifying the chemical age of the cores \citep{Pagani:09,2013A&A...551A..38P}. 
However, to fully use the capabilities of N$_2$H$^+$ as a tracer of the cloud conditions, collisional data with H$_2$ are needed.

The determination of collisional rate coefficients for N$_2$H$^+$(X$^1\Sigma^+$) with the most abundant interstellar species has been the subject of various dedicated studies in the past decades. The first realistic N$_2$H$^+$--\,He rate coefficients have been provided almost forty years ago by \cite{Green:75N2Hp}. Subsequently, \cite{Daniel:04,Daniel:05} have computed a new N$_2$H$^+$--\,He potential energy surface (PES) from highly correlated $ab$ $initio$ calculations, and derived the corresponding rotational excitation rate coefficients. By comparison with the former calculations, it was found that the differences span the range from a few percent to 100\%. The cross sections and rate coefficients between hyperfine levels were then obtained using a recoupling technique. 

However, in these two studies, the collisional partner was the He atom and recent results \citep{lique08b,Walker:14} have 
pointed out that rate coefficients for collisions with H$_{2}$($j=0$), the most abundant collisional partner in the ISM, are 
generally different from those for collisions with He. This is even more true in the case of molecular ions. Indeed, the difference 
of polarisabilities between H$_2$ and He has consequences in the long--range expansions of the corresponding PESs, which 
results in very different collisional data.
Taking into account the importance of having accurate rate coefficients for the N$_2$H$^+$--\,H$_{2}$ collisional system, we have decided to compute the first N$_2$H$^+$--\,H$_{2}$ collisional data. Such calculations are very challenging from a chemical physics point of view. Indeed, even if endothermic and inefficient at typical interstellar temperature, the N$_2$H$^+$--\,H$_{2}$ system is a reactive system. The computation of a PES for the N$_2$H$^+$--\,H$_{2}$ complex is then a difficult task that has been recently overcome by Spielfiedel et al. (in preparation). The endothermicity of the N$_2$H$^+$ + H$_2$ $\to$ N$_2$ + H$_3^+$ reaction being greater than 5000 cm$^{-1}$, purely inelastic calculations can be safely performed for collisional energies below 500 cm$^{-1}$. Hence, using this new accurate PES, we have computed state--to--state rate coefficients between the first N$_2$H$^+$ hyperfine levels and for temperatures ranging from 5~K to 70~K.  
 
The paper is organized as follows:  Section 2 describes the {\it ab initio} calculation of the PES and provides a brief description of both theory and calculations. In Sect. 3 we present and discuss our results. Section 4 is devoted to first astrophysical applications and the 
conclusions of this work are drawn in Section~\ref{sec:concl}.

\section{Methodology}

\subsection{Potential energy surface}

In our scattering calculations, we employed the recently computed N$_2$H$^+$--\,H$_2$ PES of 
Spielfiedel et al. (in preparation). 
In what follows, we briefly remind the features of the N$_2$H$^+$--\,H$_2$ PES and we refer the reader to Spielfiedel et al. (in preparation) for more details. 

The $ab$ $initio$ calculations were performed using the Jacobi coordinate system. In such a system, 
the vector $\vec{R}$ connects the N$_2$H$^+$ and H$_2$ mass centres. The rotation of the N$_2$H$^+$ 
and H$_2$ molecules is defined by the $\theta$ and $\theta'$ angles, respectively, and $\phi$ is the dihedral
angle.
The four dimension (4D) N$_2$H$^+$--\,H$_2$ PES was calculated in the supermolecular
approach, based on the single and double excitation coupled cluster method
(CCSD) \citep{Hampel92}, with perturbative contributions of
connected triple excitations computed as defined by \cite{Watts:93}
[CCSD(T)]. The calculations were performed using the augmented correlation-consistent triple zeta (aug-cc-pVTZ)
basis set \citep{dunning:89}. Both molecules were treated as rigid rotors and we fixed
the internuclear distances of H$_{2}$ and N$_2$H$^+$ at their averaged and experimental equilibrium values, respectively: 
r$_{HH}$ = 1.448 a$_{0}$ for H$_{2}$ and r$_{NN}$ = 2.065 a$_{0}$ and r$_{NH}$ = 1.955 a$_{0}$ for N$_2$H$^+$ \citep{Owrutsky:86}.

An analytical representation of the PES, noted $V(R,\theta,\theta',\phi)$, was obtained 
by expanding the interaction energies over angular functions, for distances along the $R$ coordinate,
using the following expression \citep{green:75}: 
\begin{equation} \label{eq:PES angular expansion} 
V(R,\theta,\theta',\phi) = \displaystyle\sum\limits_{l_{1}, l_{2}, l}^{} 
v_{l_{1}, l_{2}, l} (R) A_{l_{1}, l_{2}, l} (\theta, \theta ', \phi) 
\end{equation}
where $A_{l_{1}, l_{2}, l} (\theta, \theta ', \phi) $ is formed from coupled spherical functions 
associated with the rotational angular momenta of 
N$_2$H$^+$ and H$_{2}$ (see Spielfiedel et al., in preparation).
The equilibrium structure was found for a T-shape configuration with the H atom of N$_2$H$^+$ pointing towards the centre of mass of H$_{2}$. The corresponding distance between the centres of mass is 5.8 a$_{0}$ and the well depth is -2530.86 cm$^{-1}$.

Taking into account the large well depth of the PES and the relatively small rotational constant of the N$_2$H$^+$ molecule, 
we could anticipate that scattering calculations that would consider the rotational structure of both molecules would 
be prohibitive in terms of memory and CPU time.
Then, in order to simplify the calculations, we reduced the 4D PES to a 2D PES (see Spielfiedel et al., in preparation), using an adiabatic approximation \citep{Zeng:11}. In such PES, the rotation of the H$_2$ molecule is neglected. The resulting PES was fitted by means of the procedure described by \cite{Werner89} for the CN\,--\,He system. Note that the 2D PES is specifically tailored for the rotational excitation of N$_2$H$^+$ by H$_{2}(j=0)$ and cannot be used to simulate H$_{2}(j>0)$ collisions.

\begin{figure}
\begin{center}
\includegraphics[width=8.0cm,angle=0.]{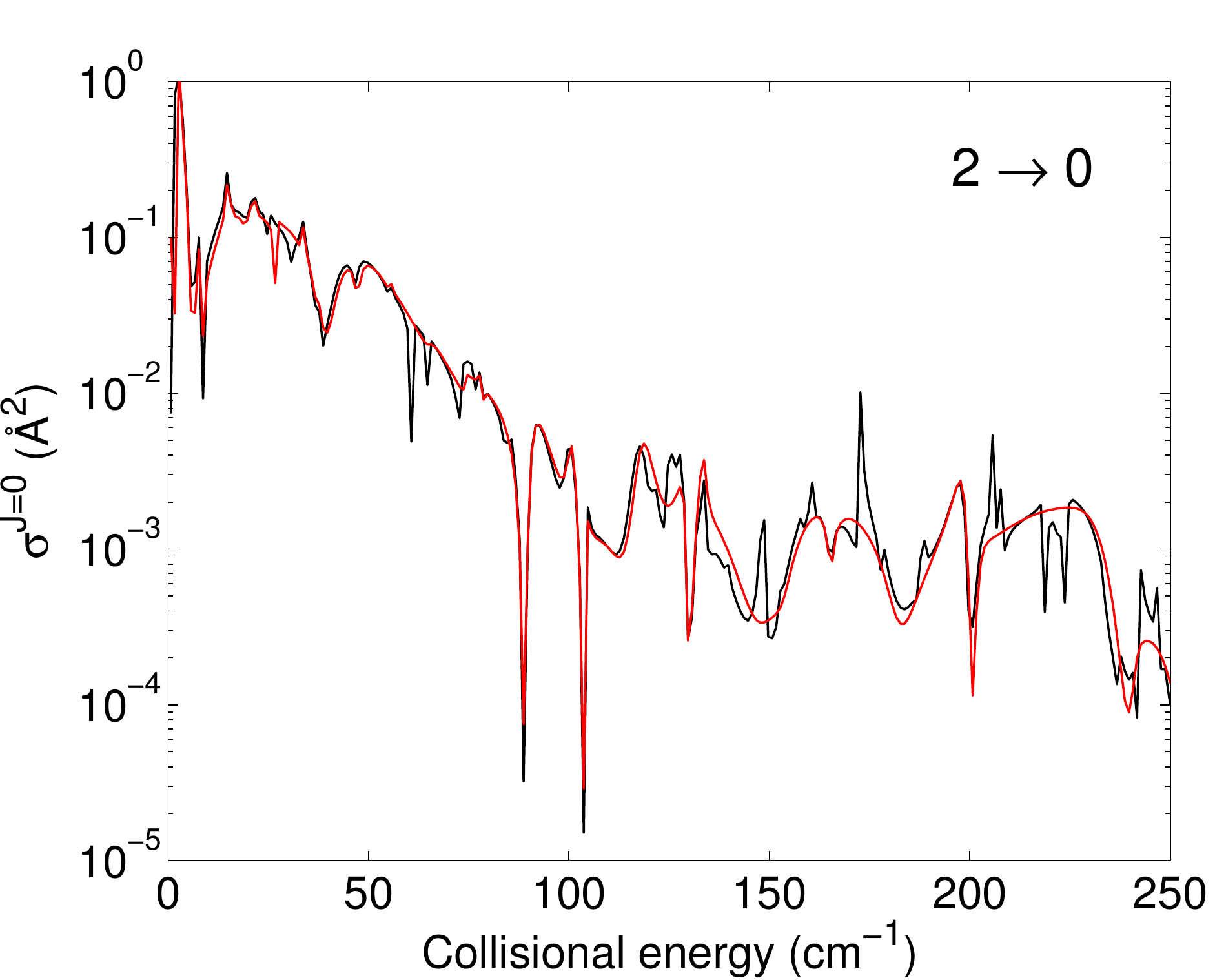}
\includegraphics[width=8.0cm,angle=0.]{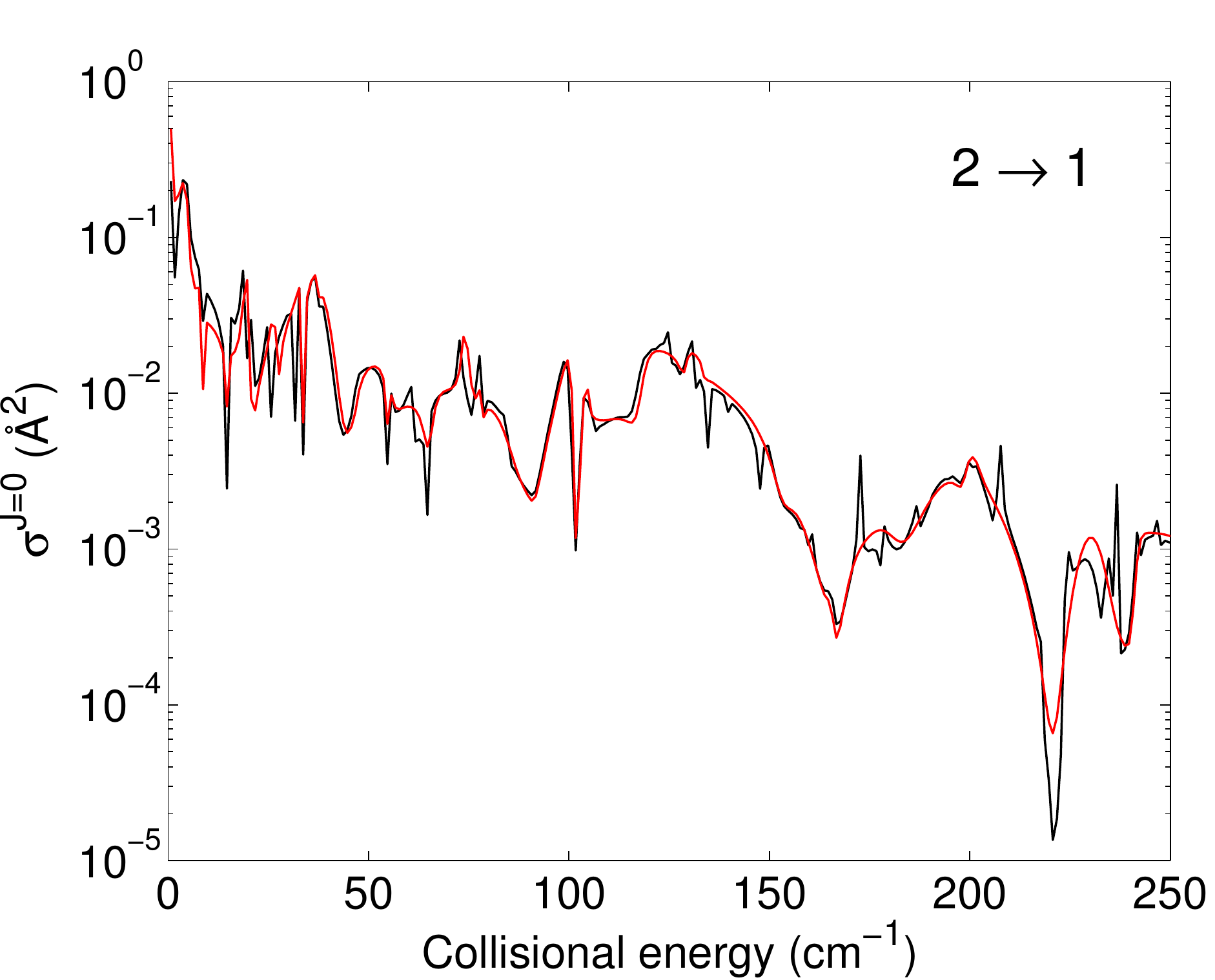}
\caption{Partial cross sections (for total angular momentum $J=0$) for the N$_2$H$^+$ molecule in collision with para-H$_{2}(j=0)$ as a function of the collision energy  (1 cm$^{-1}$ = 1.4 K) obtained from (i) 2D adiabatically reduced dimensional PES (red lines) (ii) full 4D PES  and including coupling with $j=2,4$ and 6 levels of H$_{2}$ (black lines).}
\label{fig1}
\end{center}
\end{figure}

Despite of this approximation, the accuracy of the scattering calculations 
is expected to be relatively good, as already demonstrated by \cite{Scribano:12} in the case of H$_2$O\,--\,H$_2$ collisions. 
To ascertain the error introduced by the use of the 2D adiabatically reduced dimensional PES, we compare, in Fig. \ref{fig1}, two partial cross sections obtained at a fixed total angular momentum $J=0$. The first cross section is based on the full 4D PES and includes the coupling with the $j=2, 4$ and 6 levels of H$_{2}$, which are needed to fully converge the cross sections. The second cross sections are obtained from the 2D adiabatically reduced dimensional PES and in this case, the calculations were performed just including the $j=0$ state of H$_2$. In both cases, we just considered the rotational structure of N$_2$H$^+$  and did not include the hyperfine structure. 
As it can be seen, the agreement between the two sets of data is extremely good, with differences of less 
than 5--10\% between the two sets. The main differences are the presence of resonances in the calculations based 
on the 4D PES, that are not present or well reproduced when using the 2D PES. However, resonances have been 
shown to influence only moderately the magnitude of the rate coefficients and consequently, we believe that an accurate 
description of the scattering of N$_2$H$^+$ with para--H$_{2}$ can be obtained by the simplified treatment described above.

\subsection{Scattering Calculations}

The main goal of this work is to
determine hyperfine resolved integral cross sections and
rate coefficients of  $^{14}$N$^{14}$NH$^+$ molecules induced by collisions with H$_{2}(j=0)$. Since the
rotational structure of H$_2$ is neglected, the problem, in terms
of scattering calculations, is equivalent to the collisional excitation of N$_2$H$^+$ by a structureless atom.

Since the nitrogen atoms possess a non-zero nuclear spin ($I=1$), the N$_2$H$^+$ rotational energy levels
are split in hyperfine levels which are characterized by the quantum numbers $j$, $F_1$ and $F$. Here, $F_1$
results from the coupling of $\vec{j}$ with $\vec{I_1}$ ($\vec{F_1} =
\vec{j} + \vec{I_1}$, $I_1$ being the nuclear spin of the first nitrogen atom) and $F$ results from the coupling of
$\vec{F_1}$ with $\vec{I_2}$ ($\vec{F} = \vec{F_1} + \vec{I_2}$, $I_2$ being the nuclear spin of the second nitrogen atom).
However, the hyperfine splitting of the N$_2$H$^+$ levels is very small. Hence, by assuming that the hyperfine levels are 
degenerate, it is possible to considerably simplify the hyperfine scattering problem.
Then, the integral cross sections corresponding to transitions between hyperfine levels of the N$_2$H$^+$ molecule can be obtained from scattering nuclear spin free S-matrices using a recoupling method \citep{Alexander:85hyp,Daniel:04}. 

We used \cite{Arthurs:60} description of the inelastic scattering between an atom and a linear molecule in order to obtain the close-coupling (CC) scattering matrix $S^{J}(jl;j'l')$, between rotational levels of N$_2$H$^+$. We emphasize that in the calculation of the $S^{J}(jl;j'l')$ elements, the N$_2$H$^+$ hyperfine structure is not taken into account. 
Inelastic cross sections between the hyperfine levels, from level $j,F_1,F$ to level $j',F_1',F$, are subsequently obtained through \citep{Daniel:04} :

\begin{eqnarray} \label{XSC}
\sigma_{jF_1F \to j'F_1'F'}  & = & \frac{\pi}{k^{2}_{jF_1F}} (2F_1+1)(2F_1'+1)\nonumber \\
& &  \times (2F'+1) \sum_{K}
 \left\{ 
\begin{array}{ccc}
F_1 & F_1' & K \\
F' & F & I_2 
\end{array}
\right\}^2  \nonumber \\
& & \times  \left\{ 
\begin{array}{ccc}
j & j' & K \\
F_1' & F_1 & I_1 
\end{array}
\right\}^2
P^K(j \to j')
\end{eqnarray}
The $P^{K}(j \to j')$ are the opacity tensors defined by  :
\begin{equation} \label{tens}
P^{K}(j \to j')=\frac{1}{2K+1}\sum_{ll'}|T^{K}(jl;j'l')|^{2}
\end{equation}
The reduced T-matrix elements (where $T = 1 - S$) are defined by \cite{alexander:83}: 
\begin{eqnarray}
T^{K}(jl;j'l') & = & (-1)^{-j-l'}(2K+1)\sum_{J}(-1)^{J}(2J+1) \nonumber\\
& & \times  \left\{\begin{array}{ccc}
l' & j' & J \\ 
j & l & K 
\end{array}\right\}
T^{J}(jl;j'l')
\end{eqnarray}

\subsection{Computational details}

The nuclear--spin--free $S^{J}(jl;j'l')$ matrix elements were calculated using the MOLSCAT program~\citep{molscat:94}. All calculations were made using the rigid rotor approximation. The  $^{14}$N$^{14}$NH$^+$ energy levels were computed using the rotational constants of \cite{Sastry:81}. Calculations were carried out for total energies up to 500~cm$^{-1}$. Parameters of the integrator were tested and adjusted to ensure a typical precision to within 0.05 \AA$^2$ for the inelastic cross sections and to within 1 \AA$^2$ for the elastic cross sections. For example, at total energies of 100 and 500 cm$^{-1}$, total angular momentum $J$ up to 68 and 94 were taken into account in the scattering calculations, respectively. At each energy, channels with $j$ up to 28 were included in the rotational basis to converge the calculations for all the transitions including N$_2$H$^+$ levels up to $j = 7$. Using the recoupling technique and the stored $S^{J}(jl;j'l')$ matrix elements, the opacity tensors (Eq. \ref{tens}) and the hyperfine--state--resolved cross sections (Eq. \ref{XSC}) were obtained for all hyperfine levels up to $j=7$. 

From the calculated cross sections, one can obtain the corresponding thermal rate coefficients at temperature $T$ by an average over the collision energy ($E_c$):

\begin{eqnarray}
\label{thermal_average}
k_{\alpha \rightarrow \beta}(T) & = & \left(\frac{8}{\pi\mu k^3_{B} T^3}\right)^{\frac{1}{2}}  \nonumber\\
&  & \times  \int_{0}^{\infty} \sigma_{\alpha \rightarrow \beta}\, E_{c}\, e^{-\frac{E_c}{k_{B}T}}\, dE_{c}
\end{eqnarray}
where $\sigma_{\alpha \to \beta}$ is the cross section from initial level $\alpha$ to final level $\beta$, $\mu$ is the reduced mass of the system and $k_{B}$ is Boltzmann's constant.

\section{Results}

\subsection{Hyperfine rate coefficients}

Using the computational scheme described above, we have obtained inelastic cross sections for transitions between 
the first 64 hyperfine levels of N$_2$H$^+$. With calculations performed up to a total energy of 500 cm$^{-1}$, we can 
determine the corresponding rate coefficients for temperatures up to 70 K. The complete set of (de-)excitation rate coefficients with 
$j,\,j'\le 7$ will be made available through the LAMDA \citep{schoier:05} and BASECOL \citep{Dubernet:13} databases.

\begin{figure}
\begin{center}
\includegraphics[width=8.0cm,angle=0.]{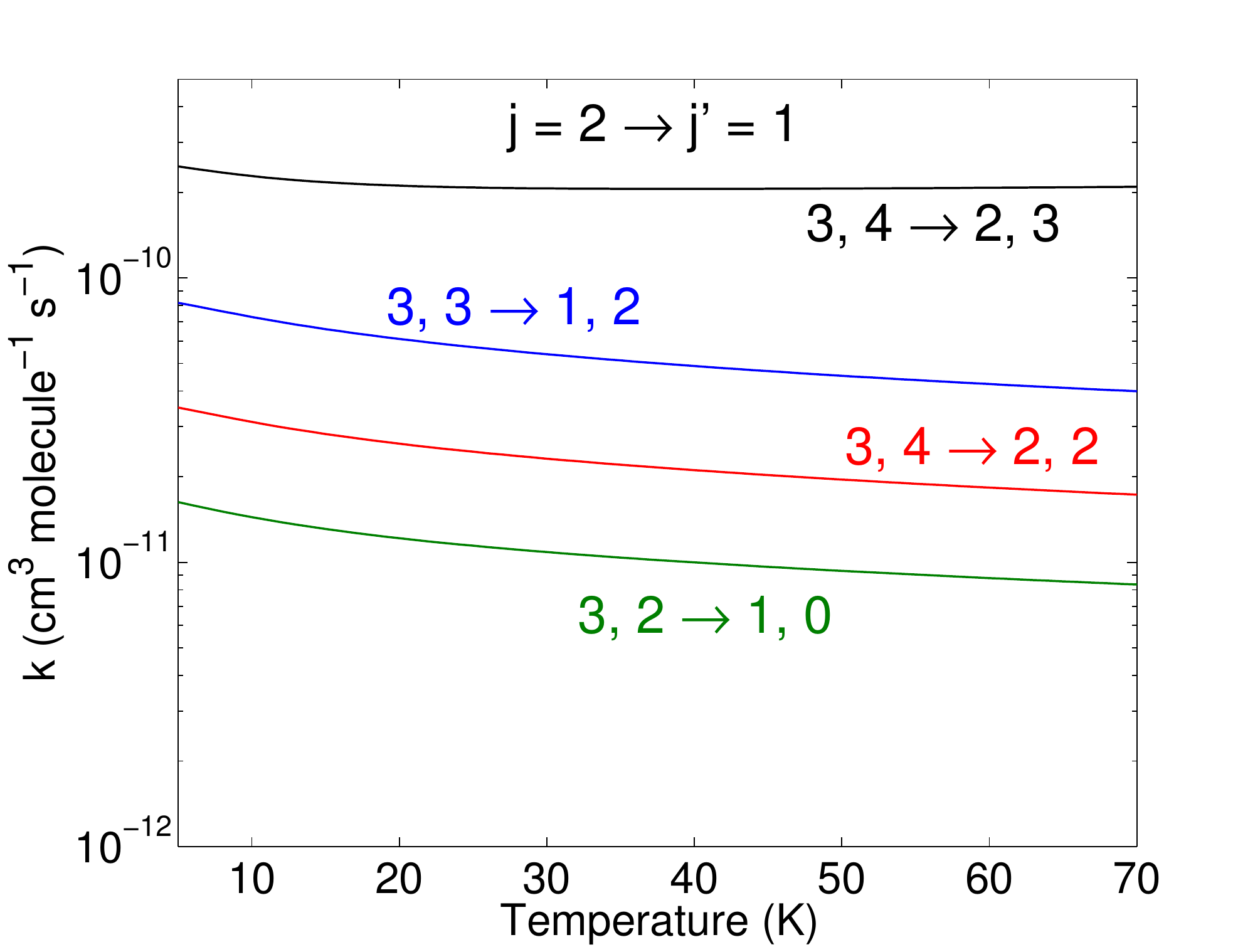}
\includegraphics[width=8.0cm,angle=0.]{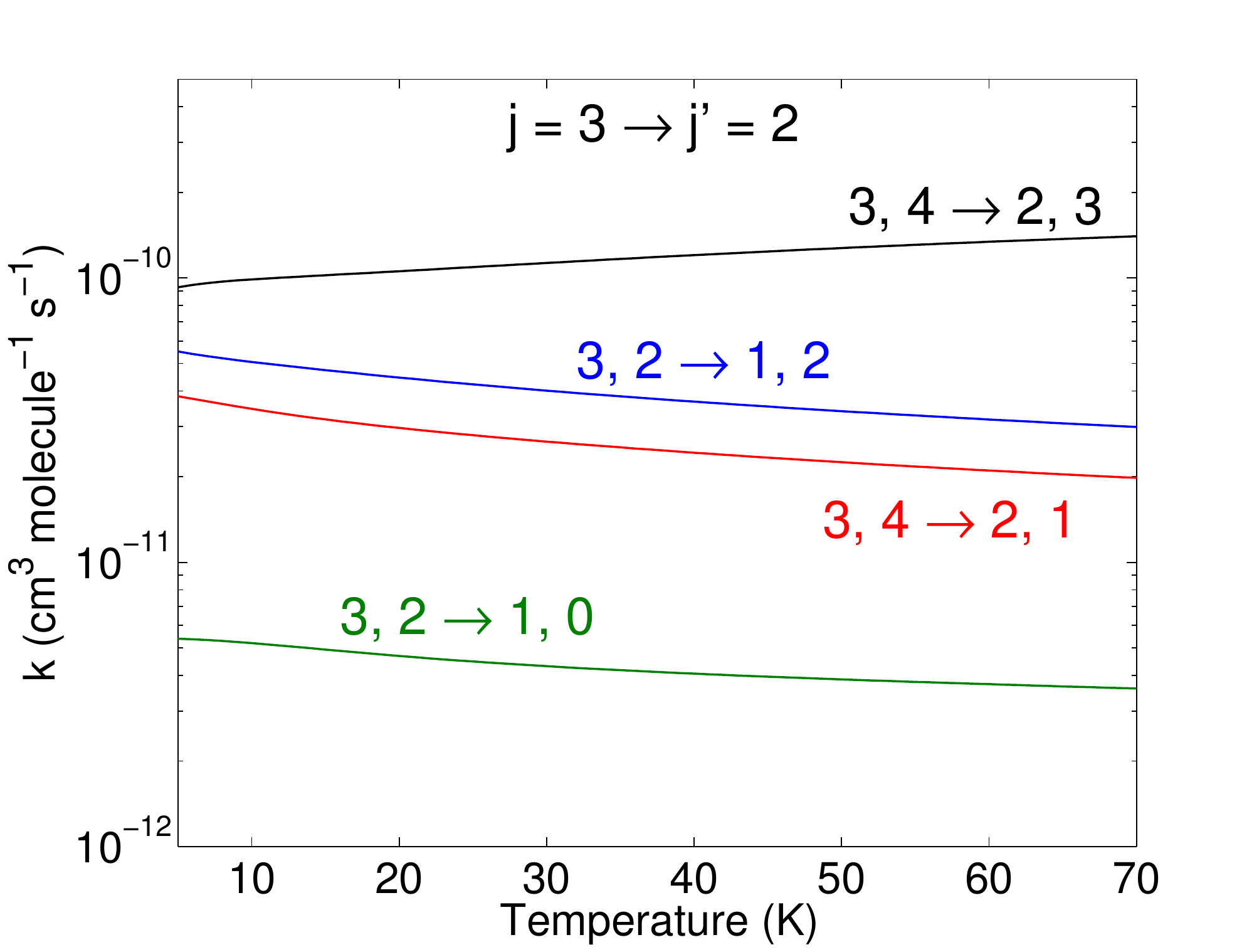}
\caption{Temperature variation of the hyperfine resolved
N$_2$H$^+$--H$_2$(j=0) rate coefficients for $j=2,F_1,F \to j'=1,F_1',F'$ and $j=3,F_1,F \to j'=2,F_1',F'$
transitions.} \label{fig2}
\end{center}
\end{figure}

Figure \ref{fig2} presents the temperature variation of the
N$_2$H$^+$--\,H$_2(j=0)$ rate coefficients for selected $j=2,F_1,F \to
j'=1,F_1',F'$ and $j=3,F_1,F \to
j'=2,F_1',F'$ transitions. 
First, one can note that the rate coefficients are weakly dependent of the temperature. This weak temperature dependance of the rate coefficients could have been anticipated, on the basis of Langevin theory for ion-neutral interactions. 
One can also note the relatively large magnitude of the rate coefficients ($k(T) > 10^{-10}$ cm$^{3}$ mol$^{-1}$ s$^{-1}$ for the dominant transitions). Usually, rate coefficients for neutral molecules such as for the isoeletronic CO molecule are of the order of magnitude of $ k(T) \simeq 10^{-11}$ cm$^{3}$ mol$^{-1}$ s$^{-1}$. The high magnitude of the rate coefficients can be related to the large well depth of the PES. This well depth is almost an order of magnitude larger than in the case of neutral molecules interacting with H$_2$. Such effect has been already observed for the HCO$^+$ cation \citep{Yazidi:14} or for the the CN$^-$ anion \citep{Klos:11} but is absent when using He as a model for H$_2$.
Hence, considering specifically H$_2$ will be of crucial importance for interstellar ions by comparison to neutral species. 

In the current case, the determination of the hyperfine propensity rules is more complex 
than for the case of only one nuclear spin, where the usual $\Delta j= \Delta F$ propensity 
is observed \citep{benabdallah:12,kalugina12,Faure:12HFS,Dumouchel:12NH}. 
In fact, as explained in \cite{Daniel:04}, because of the behaviour of the Wigner--6j coefficients in Eq. \ref{XSC},
we expect that the largest rate coefficients will satisfy (i) if $\Delta F_1 = \Delta j$ then $\Delta F =
\Delta F_1$, (ii) if $\Delta F_1 = \Delta j \pm 1$ then $\Delta F =
\Delta F_1 \pm 1$.  Such propensity rules were indeed found to describe, on the average, the 
two nuclear spins case and for the of N$_2$H$^+$--\,He system \citep{Daniel:04}. 
However, the only well defined propensity rule is $\Delta F = \Delta F_1 = \Delta j$
confirming that hyperfine rate coefficients cannot be accurately
estimated from fine structure rate coefficients using for example the
$M_j$ randomizing limit \citep{alexander:85} approach. 

\begin{figure}
\begin{center}
\includegraphics[width=8.0cm,angle=0.]{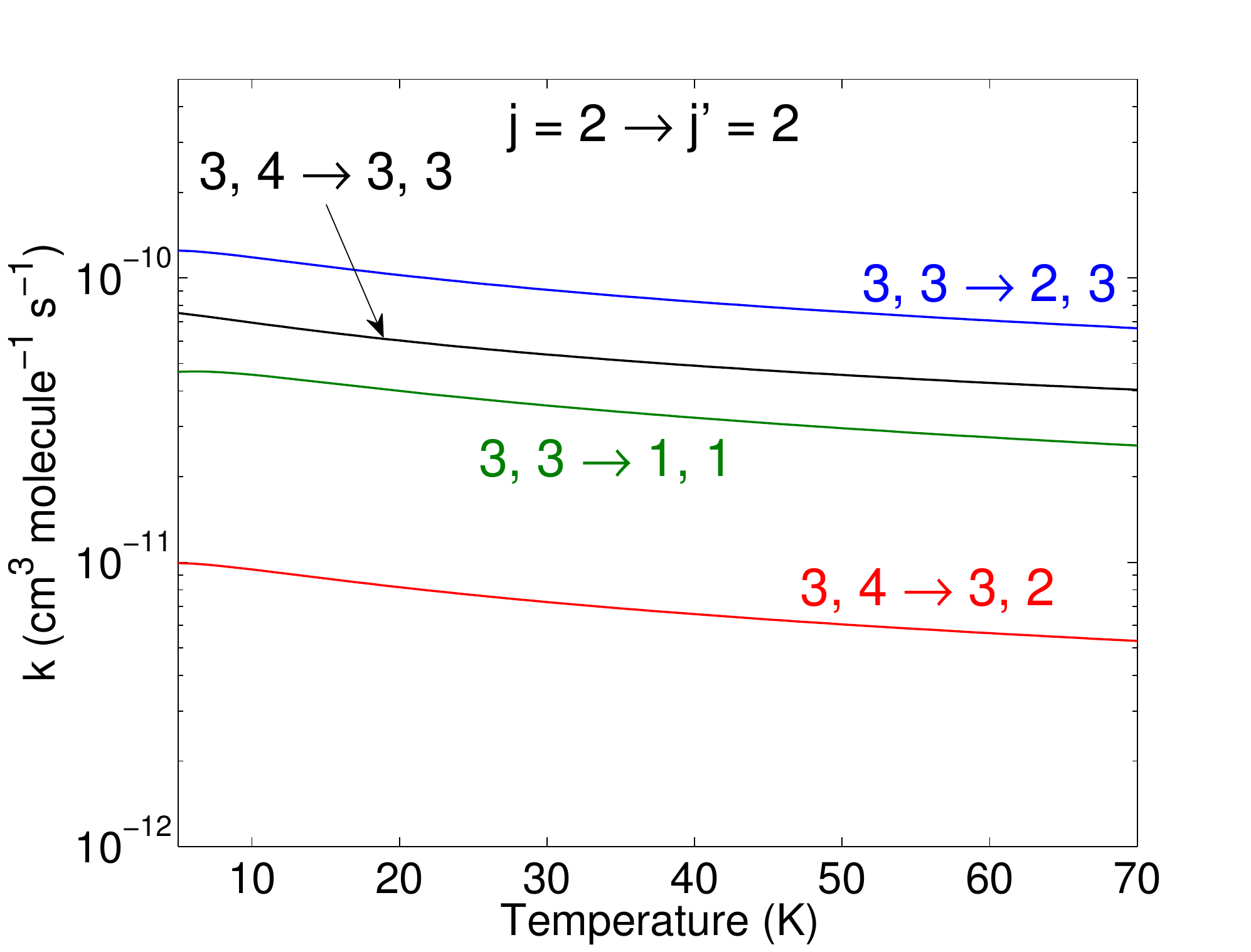}
\caption{Temperature variation of the hyperfine resolved N$_2$H$^+$--H$_2(j=0)$ rate coefficients for $j=2,F_1,F \to j'=2,F_1,F'$ transitions.}
\label{fig3}
\end{center}
\end{figure}

Finally, Figure \ref{fig3} presents the temperature variation in N$_2$H$^+$--H$_2(j=0)$ rate coefficients for the ``quasi-elastic" $j=2,F_1,F \to j'=2,F_1,F'$ transitions.
One can notice that the order of magnitude of these
``quasi-elastic" rate coefficients is similar to the pure inelastic rate coefficients. Additionally, for
transitions inside a rotational level $j$, it is difficult to extract a clear propensity rule.

\subsection{Comparison with He rate coefficients} \label{comparaison_He}

It is interesting to compare the present rate coefficients with those calculated for N$_2$H$^+$--\,He collisions \citep{Daniel:05}. Indeed, collisions with He are often used to model collisions with para-H$_{2}(j=0)$ and it is generally assumed that rate coefficients with para-H$_{2}(j=0)$ should be larger than He rate coefficients, owing to the smaller collisional reduced mass, and that the scaling factor should be $\sim$1.4 \citep{schoier:05}. We have compared in Table \ref{table:rates}, on a small sample, the N$_2$H$^+$--\,H$_{2}(j=0)$ and N$_2$H$^+$--\,He rate coefficients at 10, 30 and 50 K.
\begin{table*}
\small
  \begin{center}
\caption{Selective comparison between hyperfine rate coefficients at 10, 30 and 50 K, for collisions of N$_2$H$^+$ with para-H$_{2}(j=0)$ (this work) and those of D05 \citep{Daniel:05} for collisions of N$_2$H$^+$ with He. The rates are in units of 10$^{-10}$ cm$^{3}$ mol$^{-1}$ s$^{-1}$.}
\label{table:rates}
    \begin{tabular}{ccccccc}
\hline
\hline 
& \multicolumn{2}{c}{10K} &   \multicolumn{2}{c}{30K} &  \multicolumn{2}{c}{50K} \\
\hline 
$j,F_1,F$ $\to$ $j',F'_1,F'$ &  This work & D05 &  This work & D05 &  This work & D05 \\
1, 1, 1  $\to$ 0, 1, 0 & 0.906 & 0.384 & 0.730 & 0.325 & 0.693 & 0.307 \\
1, 2, 2  $\to$ 0, 1, 2 & 0.679 & 0.288 & 0.548 & 0.243 & 0.520 & 0.231 \\
1, 2, 3  $\to$ 0, 1, 2 & 2.716 & 1.153 & 2.191 & 0.975 & 2.081 & 0.923 \\

2, 1, 0  $\to$ 0, 1, 2 & 1.442 & 0.626 & 1.109 & 0.436 & 0.936 & 0.354 \\
2, 1, 0  $\to$ 1, 2, 1 & 0.039 & 0.016 & 0.039 & 0.016 & 0.042 & 0.018 \\

3, 2, 3  $\to$ 0, 1, 2 &  0.271 & 0.103 & 0.207 & 0.078 & 0.176 & 0.066 \\
3, 2, 3  $\to$ 1, 0, 1 &  0.411 & 0.173 & 0.340 & 0.145 & 0.311 & 0.133 \\
3, 2, 3  $\to$ 2, 1, 2 &  0.814 & 0.454 & 0.974 & 0.470 & 1.126 & 0.524 \\
3, 2, 3  $\to$ 2, 3, 4 &  0.456 & 0.154 & 0.357 & 0.120 & 0.304 & 0.105 \\

4, 3, 2  $\to$ 0, 1, 2 &  0.736 & 0.231 & 0.566 & 0.179 & 0.484 & 0.159 \\
4, 3, 2  $\to$ 2, 1, 0 &  0.190 & 0.070 & 0.182 & 0.075 & 0.175 & 0.077 \\
4, 3, 2  $\to$ 3, 2, 1 &  0.559 & 0.226 & 0.687 & 0.322 & 0.837 & 0.405 \\
4, 3, 2  $\to$ 3, 3, 3 &  0.556 & 0.198 & 0.479 & 0.182 & 0.418 & 0.167 \\
\hline
\end{tabular}
\end{center}
\end{table*}
As one can see, significant differences exist between the present and the previous results. The present data differ by up to a factor of $\sim$ 3 from the previous results, the new results being systematically larger than the previous ones. Hence, the scaling factor is clearly different from 1.4 and the ratio varies with both the temperature and the transition considered. 

Low collisional energies cross sections are very sensitive to the shape and depth of the PES well. It is then not surprising to see significant differences between the two collisional systems at low temperature. 
It could have been anticipated that scattering studies involving He as the perturber and as a representative of molecular H$_2$ are questionable in the case of molecular ions. As previously discussed, this is a consequence of the difference of 
polarizabilities with either H$_2$ or He. 

\section{Astrophysical applications}
\subsection{Model}
As stated in Sect. \ref{comparaison_He}, the current H$_2$ rate coefficients differ significantly from
the previously available rate coefficients from \citet{Daniel:05}, that considered He as a collisional partner. 
Since these rate coefficients were the only available data until now, they were used in many astrophysical 
studies. Most of the time, a scaling factor of 1.4 was used to emulate the H$_2$ rate coefficients. As discussed 
in the previous section, this factor leads to an underestimate of the actual values of the H$_2$ rate coefficients.
More recently, \citet{Bizzocchi:13} used a scaling relationship based on the HCO$^+$--\,H$_2$ and 
HCO$^+$--\,He rate coefficients. The correction applied thus ranged from factors of 1.4 and 3.2 depending 
on the transition considered, with a ratio of $\sim$2.3 applied to the transitions with $\Delta j = 1$ and $j \leq 4$.
 Given the current results, this scaling procedure gives a better estimate of the H$_2$ rate coefficients than simply
 assuming a constant 1.4 value.

\begin{figure*}
\begin{center}
\includegraphics[scale=0.5, angle=270.]{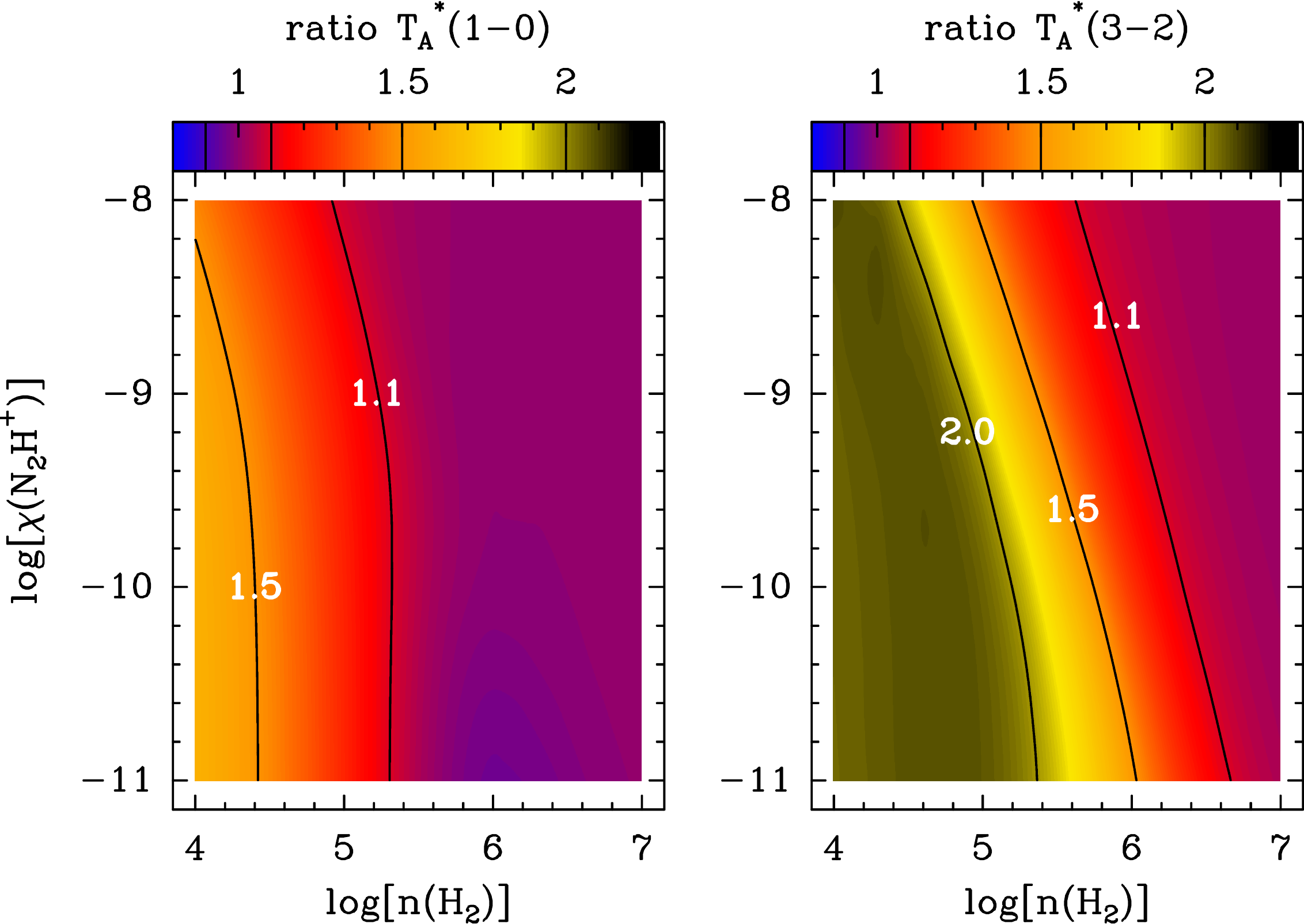}
\caption{Ratio of the $j=1-0$ (left panel) and $j=3-2$ (right panel) 
integrated intensities obtained with the H$_2$ and He rate coefficients. These ratios are given 
for a uniform spherical cloud at $T = 10$ K and by varying both the H$_2$ density and 
N$_2$H$^+$ abundance.}
\label{fig4}
\end{center}
\end{figure*}

 In order to highlight the impact of the new rate coefficients on the interpretation of the N$_2$H$^+$ observations,
 we performed a few radiative transfer calculations for physical parameters (i.e. gas density, temperature, ...) 
 and N$_2$H$^+$ abundances typical of dark clouds conditions. These calculations are based on the numerical 
 code described in \citet{Daniel:08}: the radiative transfer is solved exactly and includes the line overlap between
 the hyperfine transitions. Additionally, the N$_2$H$^+$ energy structure and line frequencies are obtained from the 
 spectroscopical constants of \citet{Caselli:95} and \citet{Pagani:09}. The line strengths and Einstein coefficients
 are obtained from the CDMS database \citep{cdms}.
 In Fig. \ref{fig4}, we compare the line intensities, integrated over frequency, for the $j=1-0$ and $3-2$ lines,
 and for radiative transfer calculations based either on the He or H$_2$ rate coefficients.   For the purpose of the comparison, we made the choice to use the antenna temperature 
scale, noted $T_A^*$.
 In the figure, we give the ratio of the results based on the H$_2$ rate coefficients over the results obtained with
 the He ones, scaled by a factor 1.4. In order to perform the calculations, we assumed a static and homogeneous spherical cloud 
 of diameter 3' at a distance of 100 pc. We fixed the gas temperature at $T = 10$ K and then varied the H$_2$ density 
 in the range $10^4 <$ $n$(H$_2$) $< 10^7$ cm$^{-3}$, and the N$_2$H$^+$ abundance in the range $10^{-11} <$ $\chi$(N$_2$H$^+$) $< 10^{-8}$ . 

\begin{figure}
\begin{center}
\includegraphics[scale=0.5, angle=270.]{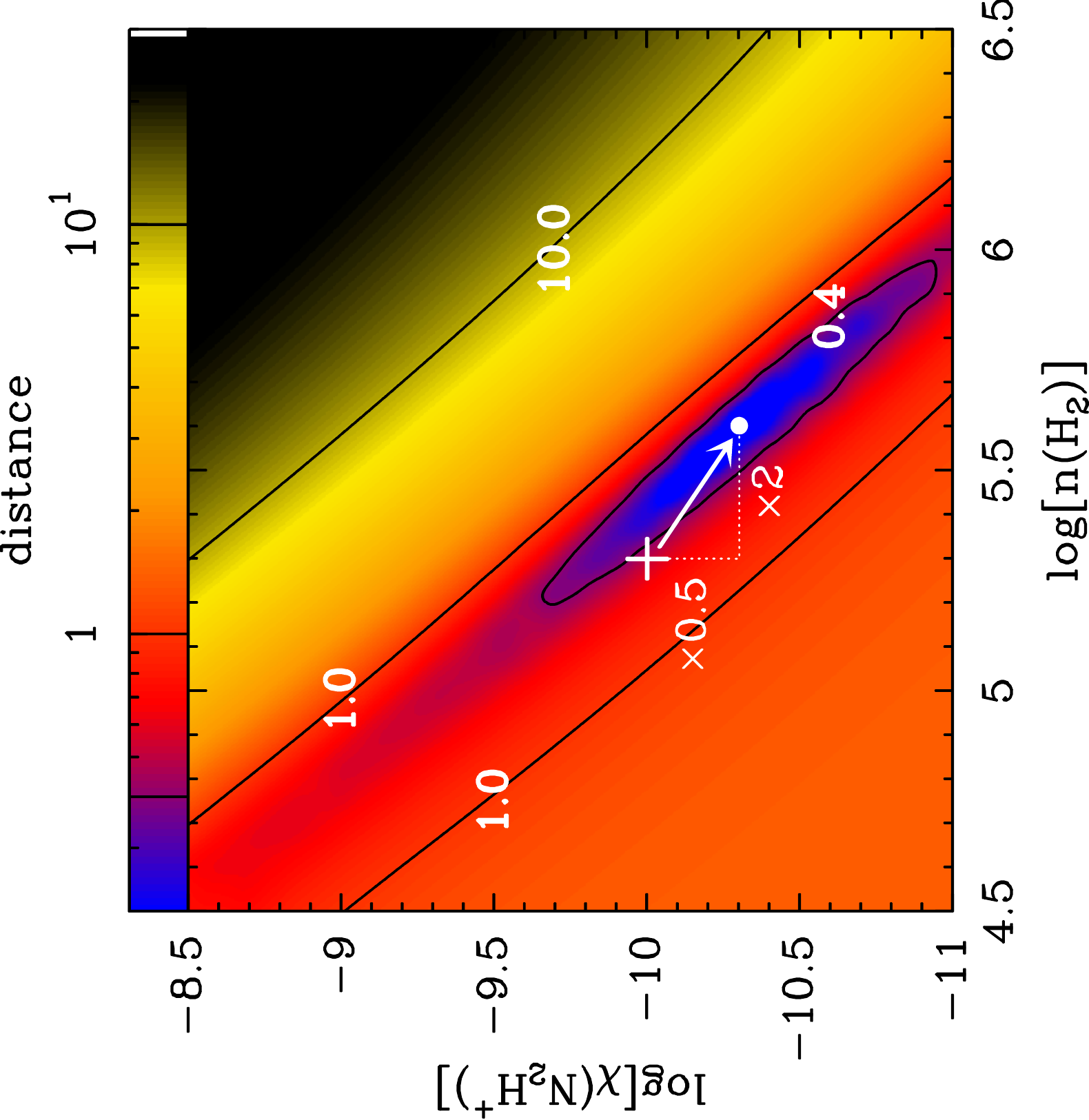}
\caption{Comparison of the intensities predicted with the H$_2$ and He rate coefficients.
The comparison is performed by mean of the distance $d$ defined in the text. By considering 
a reference model computed with the H$_2$ rate coefficients and with
$n$(H$_2$) = $2 \, 10^5$ cm$^{-3}$ and $\chi$(N$_2$H$^+$) = $10^{-10}$, we represent in this 
figure the optimum values for both $n$(H$_2$) and $\chi$(N$_2$H$^+$), 
i.e. the minimum of $d$, which enables to reproduce the reference model with the He rate coefficients.}
\label{fig5}
\end{center}
\end{figure}

From this figure, it appears that between the two sets of rate coefficients, the line intensities can show variations
as large as a factor $\sim$2. As discussed in Sect. \ref{comparaison_He}, this factor is similar to the difference that is 
obtained on average between the two sets of rate coefficients. Note that the differences are the largest at low N$_2$H$^+$
column densities (i.e. low $n$(H$_2$) and/or $\chi$(N$_2$H$^+$)), since for the highest values, 
both the $j=1-0$ and $3-2$ lines tend to thermalize. As an example, considering the $j=1-0$ line, we see in Fig. \ref{fig4}
that the integrated intensity ratio is $\sim$1 for densities $n$(H$_2$) $\geq$ $2 \, 10^5$ cm$^{-3}$.   
Finally, it can be seen in Fig. \ref{fig4} that there is a differential effect introduced by the rate coefficients
on the $j=1-0$ and $3-2$ line intensities. Indeed, over the conditions spanned by the current grid of models,
we can see that the effect is systematically larger for the $j=3-2$ line by comparison to the $j=1-0$ line.
In practice, this differential effect implies that in any astrophysical application, switching from the He rate 
coefficients to the H$_2$ rate coefficients will require to modify the estimate of both the H$_2$ density 
and the N$_2$H$^+$ abundance, and not solely one of these parameters.
In order to emphasize this point, we considered the 
$j=1-0$ and $3-2$ line intensities obtained with the H$_2$ rate coefficients and for the parameters
$n$(H$_2$) = 2\,$\times$\,10$^5$ cm$^{-3}$ and $\chi$(N$_2$H$^+$) = 1 \pdix{-10}. Subsequently, by using 
the He rate coefficients, we determined the value of $n$(H$_2$) and $\chi$(N$_2$H$^+$) which enable
to obtain similar line intensities for these two radiative transitions. The value of the H$_2$ density 
and N$_2$H$^+$ abundance are determined by minimizing the quantity  :
\begin{eqnarray}
d = 
\sqrt{
\left(\frac{T_{\textrm{He}}(1-0)}{T_{\textrm{H}_2}(1-0)}-1\right)^2
+
\left(\frac{T_{\textrm{He}}(3-2)}{T_{\textrm{H}_2}(3-2)}-1\right)^2
}
\end{eqnarray}
where $T_{\textrm{H}_2}$ and $T_{\textrm{He}}$ stands for the integrated 
intensities obtained respectively with the H$_2$ and scaled He rate coefficients.
In Fig. \ref{fig5}, the value of the distance $d$ is reported as a function of 
$n$(H$_2$) and $\chi$(N$_2$H$^+$). It appears that the $j=1-0$ and $3-2$ line intensities 
obtained with the H$_2$ rate coefficients can be reproduced with the He rate 
coefficients by varying both $n$(H$_2$) and $\chi$(N$_2$H$^+$) by factors of 2.
Indeed, the minimum of $d$ is obtained for $n$(H$_2$) = $4 \, 10^5$ cm$^{-3}$ 
and $\chi$(N$_2$H$^+$) = $5 \, 10^{-11}$. 

Hence, using the He rate coefficients 
will induce an overestimation of the H$_2$ density and an underestimation 
of the N$_2$H$^+$ abundance. Given the variations in magnitude of the rate coefficients 
between the two sets of rate coefficients, 
we can expect that a factor $\sim$2 will be typical of the modifications 
induced by the new set of rates on $n$(H$_2$) and $\chi$(N$_2$H$^+$).

\subsection{LDN 183}

Since kinetic temperature has also an impact on the excitation conditions but has to be kept realistic, we  revisited a well-known and modelled case, LDN 183, which was previously analysed with the hyperfine collisional rates computed with He \citep{2007A&A...467..179P}. The goal is to estimate the magnitude of the changes in a real case. We did not attempt to recompute the best possible fit to the data but only a plausible fit by eye to gauge the magnitude of the changes. With the former core properties as listed in \cite{2007A&A...467..179P}, Table 2 (best model), all the modelled lines are stronger than observed with the new collision rates, as expected from the model runs described above. We tried to vary separately the kinetic temperature, the density profile and the N$_2$H$^+$ abundance and found no convincing solution, again confirming the results discussed above. A workable solution consisted in lowering the temperature in the core, replacing the 7\,K constant value by a gradient from 6 to 7\,K, lowering the density by 20\,\% in the layers at temperature at 6.5\,K or above and increasing the N$_2$H$^+$ abundance by a factor 2 in the central two layers. Globally, the N$_2$H$^+$ column density decreases by a factor 8\,\% (inner core) to 20\,\% for the outer core. The N$_2$H$^+$ $j=1-0$ line being quite optically thick in LDN 183 (6 out of the 7 distinguishable hyperfine components have identical intensity towards the central position, instead of the 3:5:7 ratio observed in optically thin cases), the sensitivity of the results to the new collisional coefficients is not as dramatic as one can expect in an optically thin case but the model is more accurate and the temperature is a new low record (a temperature of 5.5 K has been advocated in the case of LDN 1544 from NH$_3$ interferometric observations, \citealt{Caselli:2007ge}).  This temperature variation impacts also the dust temperature estimate (dust and gas are thermally coupled at n(H$_2$) = 1 \pdix{6} \cmc) and consequently the dust properties.
 
\section{Conclusion}\label{sec:concl}

We have used quantum scattering calculations to investigate rotational energy transfer for collisions between the N$_2$H$^+$ and para-H$_{2}(j=0)$ molecules. The calculations were based on a new 4D potential energy surface that is then transform into a 2D adiabatically reduced dimensional PES. Rate coefficients for transitions involving the lowest hyperfine levels of the N$_2$H$^+$ molecule were determined for  temperatures ranging from 5 to 70~K. The $\Delta j = \Delta F_1 = \Delta F$ propensity rule is found for these hyperfine transitions. 

The comparison of the new  N$_2$H$^+$--\,H$_{2}$ rate coefficients  with previously calculated N$_2$H$^+$--\,He  rate coefficients shows that significant differences exist. In particular, the new H$_2$ rate coefficients are systematically larger than the rates with He, the scaling factor being around $\sim$ 3.

The consequences for astrophysical models were also evaluated. We have performed a new set of radiative transfer calculations for physical parameters and N$_2$H$^+$ abundances typical of dark cloud conditions. The line intensities of the simulated spectra,
 obtained using the new rate coefficients, are significantly larger than the ones predicted from the He rate coefficients; the factor could be as large as a factor $\sim$2 and in particular, it depends on the radiative transition. Additionally, a test on a real case (LDN 183) shows that the new rate coefficients will lead to reconsideration of the estimate of all the three: the H$_2$ density, the gas kinetic temperature  and the N$_2$H$^+$ abundance in cold molecular clouds.

\section*{Acknowledgments}
We acknowledge Pierre Hily Blant and Alexandre Faure for fruitful discussions.
This research was supported by the CNRS national program  ``Physique et Chimie du Milieu Interstellaire''. FL and FD also thank the Agence Nationale de la Recherche (ANR-HYDRIDES), contract ANR-12-BS05-0011-01 and the CPER
Haute-Normandie/CNRT/Energie, Electronique, Mat\'eriaux.

\bibliographystyle{mn2e}

%\bibliography{vanderwaals}

\begin{thebibliography}{46}
\expandafter\ifx\csname natexlab\endcsname\relax\def\natexlab#1{#1}\fi

\bibitem[{Alexander(1985)}]{alexander:85}
Alexander M.~H., 1985, Chem. Phys., 92, 337

\bibitem[{{Alexander} \& {Dagdigian}(1983)}]{alexander:83}
{Alexander} M.~H., {Dagdigian} P.~J., 1983, \jcp, 79, 302

\bibitem[{{Alexander} \& {Dagdigian}(1985)}]{Alexander:85hyp}
{Alexander} M.~H., {Dagdigian} P.~J., 1985, \jcp, 83, 2191

\bibitem[{{Arthurs} \& {Dalgarno}(1960)}]{Arthurs:60}
{Arthurs} A.~M., {Dalgarno} A., 1960, Proc. R. Soc. A, 256, 540

\bibitem[{{Ben Abdallah} {et~al}\mbox{.}(2012){Ben Abdallah}, {Najar},
  {Jaidane}, {Dumouchel}, \& {Lique}}]{benabdallah:12}
{Ben Abdallah} D., {Najar} F., {Jaidane} N., {Dumouchel} F., {Lique} F., 2012,
  \mnras, 419, 2441

\bibitem[{{Bizzocchi} {et~al}\mbox{.}(2013){Bizzocchi}, {Caselli}, {Leonardo},
  \& {Dore}}]{Bizzocchi:13}
{Bizzocchi} L., {Caselli} P., {Leonardo} E., {Dore} L., 2013, \aap, 555, A109

\bibitem[{{Caselli} {et~al}\mbox{.}(1995){Caselli}, {Myers}, \&
  {Thaddeus}}]{Caselli:95}
{Caselli} P., {Myers} P.~C., {Thaddeus} P., 1995, \apjl, 455, L77

\bibitem[{Crapsi {et~al}\mbox{.}(2007)Crapsi, Caselli, Walmsley, \&
  Tafalla}]{Caselli:2007ge}
Crapsi A., Caselli P., Walmsley M., Tafalla M., 2007, A{\&}A, 470, 221

\bibitem[{{Daniel} \& {Cernicharo}(2008)}]{Daniel:08}
{Daniel} F., {Cernicharo} J., 2008, \aap, 488, 1237

\bibitem[{{Daniel} {et~al}\mbox{.}(2004){Daniel}, {Dubernet}, \&
  {Meuwly}}]{Daniel:04}
{Daniel} F., {Dubernet} M.-L., {Meuwly} M., 2004, \jcp, 121, 4540

\bibitem[{{Daniel} {et~al}\mbox{.}(2005){Daniel}, {Dubernet}, {Meuwly},
  {Cernicharo}, \& {Pagani}}]{Daniel:05}
{Daniel} F., {Dubernet} M.-L., {Meuwly} M., {Cernicharo} J., {Pagani} L., 2005,
  \mnras, 363, 1083

\bibitem[{{Dubernet, M.-L.} {et~al}\mbox{.}(2013){Dubernet, M.-L.}, {Alexander,
  M. H.}, {Ba, Y. A.}, {Balakrishnan, N.}, {Balan\c{c}a, C.}, {Ceccarelli, C.},
  {Cernicharo, J.}, {Daniel, F.}, {Dayou, F.}, {Doronin, M.}, {Dumouchel, F.},
  {Faure, A.}, {Feautrier, N.}, {Flower, D. R.}, {Grosjean, A.}, {Halvick, P.},
  {Klos, J.}, {Lique, F.}, {McBane, G. C.}, {Marinakis, S.}, {Moreau, N.},
  {Moszynski, R.}, {Neufeld, D. A.}, {Roueff, E.}, {Schilke, P.}, {Spielfiedel,
  A.}, {Stancil, P. C.}, {Stoecklin, T.}, {Tennyson, J.}, {Yang, B.},
  {Vasserot, A.-M.}, \& {Wiesenfeld, L.}}]{Dubernet:13}
{Dubernet, M.-L.} {et~al.}, 2013, \aap, 553, A50

\bibitem[{{Dumouchel} {et~al}\mbox{.}(2012){Dumouchel}, {K{\l}os}, {Tobo{\l}a},
  {Bacmann}, {Maret}, {Hily-Blant}, {Faure}, \& {Lique}}]{Dumouchel:12NH}
{Dumouchel} F., {K{\l}os} J., {Tobo{\l}a} R., {Bacmann} A., {Maret} S.,
  {Hily-Blant} P., {Faure} A., {Lique} F., 2012, \jcp, 137, 114306

\bibitem[{Dunning(1989)}]{dunning:89}
Dunning T.~H., 1989, \jcp, 90, 1007

\bibitem[{{Faure} \& {Lique}(2012)}]{Faure:12HFS}
{Faure} A., {Lique} F., 2012, \mnras, 425, 740

\bibitem[{{Green}(1975{\natexlab{a}})}]{green:75}
{Green} S., 1975{\natexlab{a}}, \jcp, 62, 2271

\bibitem[{{Green}(1975{\natexlab{b}})}]{Green:75N2Hp}
{Green} S., 1975{\natexlab{b}}, \apj, 201, 366

\bibitem[{{Hampel} {et~al}\mbox{.}(1992){Hampel}, {Peterson}, \&
  {Werner}}]{Hampel92}
{Hampel} C., {Peterson} K.~A., {Werner} H.-J., 1992, Chem. Phys. Lett., 190, 1

\bibitem[{{Havenith} {et~al}\mbox{.}(1990){Havenith}, {Zwart}, {Leo Meerts}, \&
  {Ter Meulen}}]{Havenith:90}
{Havenith} M., {Zwart} E., {Leo Meerts} W., {Ter Meulen} J.~J., 1990, \jcp, 93,
  8446

\bibitem[{{Hily-Blant} {et~al}\mbox{.}(2008){Hily-Blant}, {Walmsley}, {Pineau
  Des For{\^e}ts}, \& {Flower}}]{2008A&A...480L...5H}
{Hily-Blant} P., {Walmsley} M., {Pineau Des For{\^e}ts} G., {Flower} D., 2008,
  \aap, 480, L5

\bibitem[{Hutson \& Green(1994)}]{molscat:94}
Hutson J.~M., Green S., 1994. {\sc molscat} computer code, version 14 (1994),
  distributed by Collaborative Computational Project No. 6 of the Engineering
  and Physical Sciences Research Council (UK)

\bibitem[{Indriolo {et~al}\mbox{.}(2013)Indriolo, Neufeld, Seifahrt, \&
  Richter}]{Indriolo:2013cj}
Indriolo N., Neufeld D., Seifahrt A., Richter M.~J., 2013, ApJ, 764, 188

\bibitem[{{Kalugina} {et~al}\mbox{.}(2012){Kalugina}, {Lique}, \&
  {K{\l}os}}]{kalugina12}
{Kalugina} Y., {Lique} F., {K{\l}os} J., 2012, \mnras, 422, 812

\bibitem[{{Keto} \& {Caselli}(2008)}]{Caselli:2008dm}
{Keto} E., {Caselli} P., 2008, ApJ, 683, 238

\bibitem[{{K{\l}os} \& {Lique}(2011)}]{Klos:11}
{K{\l}os} J., {Lique} F., 2011, \mnras, 418, 271

\bibitem[{Lee {et~al}\mbox{.}(2001)Lee, Myers, \& Tafalla}]{Lee:2001jm}
Lee C.~W., Myers P.~C., Tafalla M., 2001, ApJS, 136, 703

\bibitem[{Lemme {et~al}\mbox{.}(1995)Lemme, Walmsley, Wilson, \&
  Muders}]{Lemme:1995wz}
Lemme C., Walmsley M., Wilson T.~L., Muders D., 1995, A{\&}A, 302, 509

\bibitem[{{Lique} {et~al}\mbox{.}(2008){Lique}, {Tobo{\l}a}, {K{\l}os},
  {Feautrier}, {Spielfiedel}, {Vincent}, {Cha{\l}asi{\'n}ski}, \&
  {Alexander}}]{lique08b}
{Lique} F., {Tobo{\l}a} R., {K{\l}os} J., {Feautrier} N., {Spielfiedel} A.,
  {Vincent} L.~F.~M., {Cha{\l}asi{\'n}ski} G., {Alexander} M.~H., 2008, \aap,
  478, 567

\bibitem[{{M{\"u}ller} {et~al}\mbox{.}(2005){M{\"u}ller}, {Schl{\"o}der},
  {Stutzki}, \& {Winnewisser}}]{cdms}
{M{\"u}ller} H.~S.~P., {Schl{\"o}der} F., {Stutzki} J., {Winnewisser} G., 2005,
  Journal of Molecular Structure, 742, 215

\bibitem[{Owrutsky {et~al}\mbox{.}(1986)Owrutsky, Guderman, Martner, Tack,
  Rosenbaum, \& Saykally}]{Owrutsky:86}
Owrutsky J.~C., Guderman C.~S., Martner C.~C., Tack L.~M., Rosenbaum N.~H.,
  Saykally R.~J., 1986, \jcp, 84, 605

\bibitem[{{Pagani} {et~al}\mbox{.}(2007){Pagani}, {Bacmann}, {Cabrit}, \&
  {Vastel}}]{2007A&A...467..179P}
{Pagani} L., {Bacmann} A., {Cabrit} S., {Vastel} C., 2007, \aap, 467, 179

\bibitem[{Pagani {et~al}\mbox{.}(2012)Pagani, Bourgoin, \&
  Lique}]{2012A&A...548L...4P}
Pagani L., Bourgoin A., Lique F., 2012, A{\&}A, 548, L4

\bibitem[{Pagani {et~al}\mbox{.}(2013)Pagani, Lesaffre, Jorfi, Honvault,
  Gonz{\'a}lez-Lezana, \& Faure}]{2013A&A...551A..38P}
Pagani L., Lesaffre P., Jorfi M., Honvault P., Gonz{\'a}lez-Lezana T., Faure
  A., 2013, A{\&}A, 551, 38

\bibitem[{Pagani {et~al}\mbox{.}(2005)Pagani, Pardo, Apponi, Bacmann, \&
  Cabrit}]{2005A&A...429..181P}
Pagani L., Pardo J.~R., Apponi A.~J., Bacmann A., Cabrit S., 2005, A{\&}A, 429,
  181

\bibitem[{{Pagani} {et~al}\mbox{.}(2009){Pagani}, {Vastel}, {Hugo},
  {Kokoouline}, {Greene}, {Bacmann}, {Bayet}, {Ceccarelli}, {Peng}, \&
  {Schlemmer}}]{Pagani:09}
{Pagani} L. {et~al.}, 2009, \aap, 494, 623

\bibitem[{{Sastry} {et~al}\mbox{.}(1981){Sastry}, {Helminger}, {Herbst}, \& {De
  Lucia}}]{Sastry:81}
{Sastry} K.~V.~L.~N., {Helminger} P., {Herbst} E., {De Lucia} F.~C., 1981,
  Chemical Physics Letters, 84, 286

\bibitem[{{Sch{\"o}ier} {et~al}\mbox{.}(2005){Sch{\"o}ier}, {van der Tak}, {van
  Dishoeck}, \& {Black}}]{schoier:05}
{Sch{\"o}ier} F.~L., {van der Tak} F.~F.~S., {van Dishoeck} E.~F., {Black}
  J.~H., 2005, \aap, 432, 369

\bibitem[{{Scribano} {et~al}\mbox{.}(2012){Scribano}, {Faure}, \&
  {Lauvergnat}}]{Scribano:12}
{Scribano} Y., {Faure} A., {Lauvergnat} D., 2012, \jcp, 136, 094109

\bibitem[{{Tafalla} {et~al}\mbox{.}(2002){Tafalla}, {Myers}, {Caselli},
  {Walmsley}, \& {Comito}}]{2002ApJ...569..815T}
{Tafalla} M., {Myers} P.~C., {Caselli} P., {Walmsley} C.~M., {Comito} C., 2002,
  \apj, 569, 815

\bibitem[{{Walker} {et~al}\mbox{.}(2014){Walker}, {Yang}, {Stancil},
  {Balakrishnan}, \& {Forrey}}]{Walker:14}
{Walker} K.~M., {Yang} B.~H., {Stancil} P.~C., {Balakrishnan} N., {Forrey}
  R.~C., 2014, ArXiv e-prints

\bibitem[{Watts {et~al}\mbox{.}(1993)Watts, Gauss, \& Bartlett}]{Watts:93}
Watts J.~D., Gauss J., Bartlett R.~J., 1993, \jcp, 98, 8718

\bibitem[{Werner {et~al}\mbox{.}(1989)Werner, Follmeg, Alexander, \&
  Lemoine}]{Werner89}
Werner H.-J., Follmeg B., Alexander M.~H., Lemoine D., 1989, \jcp, 91, 5425

\bibitem[{Willacy {et~al}\mbox{.}(1998)Willacy, Langer, \&
  Velusamy}]{Willacy:1998bl}
Willacy K., Langer W.~D., Velusamy T., 1998, ApJ, 507, L171

\bibitem[{{Yazidi} {et~al}\mbox{.}(2014){Yazidi}, {Ben Abdallah}, \&
  {Lique}}]{Yazidi:14}
{Yazidi} O., {Ben Abdallah} D., {Lique} F., 2014, \mnras, 441, 664

\bibitem[{{Zeng} {et~al}\mbox{.}(2011){Zeng}, {Li}, {Le Roy}, \&
  {Roy}}]{Zeng:11}
{Zeng} T., {Li} H., {Le Roy} R.~J., {Roy} P.-N., 2011, \jcp, 135, 094304

\end{thebibliography}

\bsp

\label{lastpage}

\end{document}